# Defect engineering of magnetic ground state in EuTiO$_3$ epitaxial thin films

Dongwon Shin[a,#], Inseo Kim[b,#], Sehwan Song[c], Yu-Seong Seo[a], Jungseek Hwang[a], Sungkyun Park[c], Minseok Choi[b,*], and Woo Seok Choi[a,*]

[a]*Department of Physics, Sungkyunkwan University, Suwon 16419, Republic of Korea*

[b]*Department of Physics, Inha University, Incheon 22212, Republic of Korea*

[c]*Department of Physics, Pusan National University, Pusan 46241, Republic of Korea*

*E-mails: minseok.choi@inha.ac.kr, choiws@skku.edu

**ABSTRACT**

Atomistic defect engineering through the pulsed laser epitaxy of perovskite transition metal oxides offers facile control of their emergent opto-electromagnetic and energy properties. Among the various perovskite oxides, $EuTiO_3$ exhibits a strong coupling between the lattice, electronic, and magnetic degrees of freedom. This coupling is highly susceptible to atomistic defects. In this study, we investigated the magnetic phase of $EuTiO_3$ epitaxial thin films via systematic defect engineering. A magnetic phase transition from an antiferromagnet to a ferromagnet was observed when the unit cell volume of $EuTiO_3$ expanded due to the introduction of Eu-O vacancies. Optical spectroscopy and density functional theory calculations show that the change in the electronic structure as the ferromagnetic phase emerges can be attributed to the weakened Eu-Ti-Eu super-exchange interaction and the developed ferromagnetic Eu-O-Eu interaction. Facile defect engineering in $EuTiO_3$ thin films facilitates understanding and tailoring of their magnetic ground state.

# 1. INTRODUCTION

The deliberate design and control of defects in transition metal oxide thin films lead to the selective enhancement of their functional behaviors. In particular, various emergent optoelectronic and magnetic properties can be tailored in complex oxides by introducing elemental vacancies, which extensively modify the crystalline lattice and electronic structures[1-4]. For example, $SrTiO_3$ (STO) thin films are prototypical perovskite oxides that exhibit defect-induced phase transitions resulting in superconductivity, metal-to-insulator transition, and ferroelectricity[5-8]. In $VO_2$ epitaxial thin films, increasing level of H doping triggers a two-step electronic phase modulation from an insulator phase ($VO_2$) to a metallic phase ($H_xVO_2$) to another insulator phase ($HVO_2$)[9]. In $BiFeO_3$[10] and $BaTiO_3$ thin films, defect dipoles due to controlled cation and oxygen vacancies induce ferroelectric polarizations[11]. In $SrRuO_3$ thin films, both Sr and Ru-O vacancies disturb the ferromagnetic (FM) ordering, whereas the Ru-O vacancies induce a structural phase transition from a tetragonal to an orthorhombic structure[12]. The crystal structural phase transition is further accompanied by a substantial modification of the Ru-O hybridization, which gives rise to the enhanced electrocatalytic activity[13]. In addition, $YMnO_3$ thin film undergoes an antiferromagnetic (AFM) to FM transition with the introduction of oxygen vacancies[14]. When the oxygen vacancy concentration is further increased, ordering of the oxygen vacancies occurs, resulting in a topotactic phase transition to a brownmillerite structure. Similar transitions occur in $SrCoO_x$ and $SrFeO_x$ thin films[15, 16], in which both electronic and magnetic phase transitions occur simultaneously[17, 18].

Perovskite $EuTiO_3$ (ETO) is a promising candidate for studying defect-related physics because the energy scales of the lattice, charge, and spin degrees of freedom are similar and can hence be tuned by multiple parameters such as epitaxial strain, isotropic pressure, chemical doping, and electric field[19-40]. Bulk ETO is a G-type AFM with a spin moment of $J = S = 7/2$ on each Eu site[19, 20]. Bulk ETO undergoes a structural phase transition at 282 K, from the cubic $Pm\bar{3}m$ to tetragonal $I4/mcm$ structure[41]. The Poisson's ratio has been assumed as 0.24[42]. Interestingly, by applying an epitaxial strain (both compressive and tensile), isotropic pressure, chemical doping (e.g., La for Eu and Nb, Al, and Gd for Ti), or an electric field, FM ordering can be induced[25-38]. A modification of the exchange coupling between the local Eu spins was attributed to the magnetic phase transition in which the AFM and FM phases were nearly degenerate. Defect engineering can also induce the FM phase[39, 40]. The itinerant $3d$ electrons at the Ti site introduced by oxygen vacancies were speculated to induce the Ruderman-Kittel-Kasuya-Yoshida (RKKY) interaction, leading to an FM state in ETO thin films grown on STO substrates. However, as the STO substrates are highly susceptible to oxygen vacancies at low pressure and high temperature, it is difficult to isolate the transport properties of the ETO thin film alone[39]. Despite the proposed various interpretations, the origin of the FM phase in ETO is still unclear. Therefore, a comprehensive study considering the crystalline lattice, electronic structure, and magnetic phase is necessary to understand the emergent FM phase in ETO.

In this study, we present the facile defect engineering of the FM phase in ETO epitaxial thin films. We fabricated epitaxial ETO thin films with modified unit cell (u.c.) volumes via defect engineering by pulsed laser epitaxy (PLE). Figure 1 schematically summarizes the emergence of the FM phase with the defect engineering-induced u.c. volume expansion in the ETO thin films. The modification of the electronic structure, consistently evidenced by optical spectroscopy measurements and density functional theory (DFT) calculations, plays a key role in the emergence of the FM phase. In particular, defect engineering results in a preferential FM interaction via Eu-O-Eu super-exchange over the conventional AFM interaction via Eu-Ti-Eu super-exchange. We propose a mechanism for the emergence of the FM phase in ETO thin films with elemental defects and provide a facile route for tailoring the magnetic ground state from the AFM to the FM phase.

## 2. EXPERIMENTAL and THEORETICAL PROCEDURES

### 2.1 Growth and post-annealing of epitaxial thin films

ETO epitaxial thin films were fabricated using PLE on LSAT ((LaAlO$_3$)$_{0.3}$(Sr$_2$AlTaO$_8$)$_{0.7}$) and STO substrates. An excimer KrF laser ($\lambda$ = 248 nm, IPEX864; Lightmachinery) with a fluence of 1.5 $J$/cm$^2$ and a repetition rate of 5 Hz was used. The thin films were synthesized at 700 ºC under various oxygen partial pressures ($P$(O$_2$) = ~3 × 10$^{-7}$ (base pressure), 10$^{-6}$, 10$^{-5}$, and 10$^{-4}$ Torr). For post-annealing, the as-grown samples were sealed in a quartz tube under an Ar environment of 10 Torr and annealed at 1000 ºC for 8.5 h.

### 2.2 Lattice structure and chemical composition characterization

The crystal structure and lattice parameters of the ETO thin films on LSAT and STO substrates were characterized by high-resolution X-ray diffraction (XRD; PANalytical X'Pert Pro). X-ray photoelectron spectroscopy (XPS; AXIS SUPRA, KRATOS analytical) with Al $K\alpha$ radiation was used to study the chemical state and composition of the ETO thin films on LSAT and STO substrates. All the X-ray photoelectron spectra were calibrated using the C-C bonding peak (284.5 eV). Eu 4$d$, Ti 2$p$, and O 1$s$ spectra of the ETO thin films were deconvoluted using Gaussian-Lorentz curves.

### 2.3 Magnetization measurements

Magnetization measurements were performed using a magnetic property measurement system (MPMS3; Quantum Design). The temperature-dependent of magnetization $M(T)$ was measured from 300 to 2 K under a magnetic field of 100 Oe along the in-plane direction of the thin films. The magnetic field dependent magnetization $M(H)$ was measured along the in-plane direction of the thin films at 2 K.

## 2.4 Optical spectroscopy

The transmittance spectra of the ETO thin films on both-side-polished LSAT were obtained using a commercial monochromatic spectrophotometer (Lambda 950; PerkinElmer) over a spectral range from mid-infrared to UV (0.6–6.2 eV). The STO substrates were not suitable for the experiment, as they were highly susceptible to the creation of oxygen vacancies and became opaque when annealed in a vacuum at high temperatures.

## 2.5 Theoretical analysis

First-principles calculations were performed using the projector augmented-wave method as implemented in the VASP code[43]. The exchange-correlation functional constructed via the generalized-gradient-approximation in the Perdew-Burke-Ernzerhof scheme[44] was used. A rotationally invariant $+U$ method[45] was applied for the Eu *4f* states with an on-site Coulomb interaction parameter $U = 5.57$ eV and Hund's exchange $J = 1.0$ eV. These parameters have been reported to reproduce the structural parameters and magnetic properties in ETO well[46]. $\sqrt{2} \times \sqrt{2} \times 2$ tetragonal supercells containing 20 atoms and $7 \times 7 \times 5$ *k*-point grids were used to examine various magnetic configurations of the stoichiometric ETO. For the defect simulation, $2\sqrt{2} \times 2\sqrt{2} \times 2$ supercells containing 80 atoms and $3 \times 3 \times 5$ *k*-point grids were used. For chemical bond analysis, the integration of the crystal orbital Hamiltonian populations (ICOHP) was examined using the LOBSTER code[47]. The ICOHP is a measure of the covalent bond strength: a large (small) value of −ICOHP indicates a strong (weak) bond strength.

## 3. RESULTS AND DISCUSSION

Figures 2 (LSAT substrates) and S1 (STO substrates) show the systematic structural modulation of the ETO epitaxial thin films via defect engineering in terms of the X-ray diffraction (XRD) $\theta$-$2\theta$ scans and reciprocal space maps (RSM). The bulk lattice parameter of cubic ETO is 3.905 Å. When epitaxially strained ETO films are grown on LSAT substrates ($a_{LSAT} = 3.868$ Å), the

*c*-axis lattice parameter increases, as shown in Fig. 2(a). Additionally, the *c*-axis lattice parameter systematically increases with the decrease in oxygen partial pressure ($P(O_2)$). This $P(O_2)$-dependence is consistently displayed in the films grown on STO substrates ($a_{STO} = 3.905$ Å) (Fig. S1(a)). While the nominal epitaxial strain induced in ETO thin films on STO substrates is zero, the actual lattice parameters systematically deviate from the nominal values resulting from defect engineering. To induce additional changes in the lattice parameters, we post-annealed the thin films in a 10 Torr Ar environment at 1000 ºC for 8.5 h. Upon post-annealing, the *c*-axis lattice parameters approach the bulk value (Fig. 2(a)) as the epitaxial strain relaxes on the LSAT substrates (Figs. 2(b)−(e)). Again, the same tendency is observed for the films grown on STO substrates (Fig. S1), although the epitaxial strain is not relaxed there. To compare the crystalline lattices of the various ETO thin films (grown and post-annealed under different conditions) on an equal footing, we extracted the change in the u.c. volume (*V*) for the ETO. Figure 2(f) summarizes the relationship between the Eu/Ti ratio obtained from the X-ray Photoelectron Spectroscopy (XPS) measurements (see Supplementary Table S1) and *V* of ETO thin films fabricated on both LSAT and STO substrates. In general (except for the two stray data points, which we ignore in future discussions), *V* increases systematically as the Eu/Ti ratio decreases. The increase in *V* might have originated from the enhanced Coulomb repulsion in the absence of Eu-O ions (both Eu and O ions)[48]. We do note that it is rather unclear how the post-annealing in highly reducing condition affects the Eu/Ti ratio in addition to the introduction of oxygen vacancies in the ETO thin films. As observed in our XPS results (see Supplementary Table S1), it seems that the Eu and/or Ti ions are also being evaporated along with the oxygen ions at the high temperature condition.

The ETO thin films with increased *V* consistently exhibit larger saturation magnetization ($M_s$) accompanied by the FM phase. The temperature-dependent magnetization, $M(T)$, of the ETO thin films grown on LSAT substrates was obtained using field-cooled cooling at a magnetic field of 100 Oe applied along the in-plane direction, as shown in Fig. 3(a). While the as-grown thin film at $P(O_2) = 10^{-4}$ Torr (green solid line) does not show any change across the experimental temperature range, the other samples commonly exhibit a clear FM transition at $T_C = \sim 5$ K. Other than the sharp increase below $T_C$, we do not find different magnetic signatures (see the inset of Fig. 3(a) for a wider temperature scan), consistent with previous reports[39, 49]. The magnetic field-dependent magnetization, $M(H)$, of the ETO thin films grown at the base pressure and the post-annealed thin films measured at 2 K (Fig. 3(b)) also demonstrates the FM phase. The magnetization shows clear hysteretic behaviors (see the inset of Fig. 3(b)) with a large $M_s$ value approaching $\sim 7$ $\mu_B$/Eu and a finite coercive field ($H_c$) of 9.6 Oe[39, 40, 49, 50]. On the other hand, the as-grown thin film at $P(O_2) = 10^{-4}$ Torr shows highly suppressed magnetization with a $M_s$ value of $\sim 1.3$ $\mu_B$/Eu and a $H_c$ of 0 Oe. Again, very similar evolutions of the magnetic behavior with the systematic modification of the growth parameters and post-annealing were also observed for the ETO thin films grown on STO substrates (Fig. S2). Notably, the emergence of the FM phase in ETO thin films can be primarily correlated to $V$. ETO thin films with large $V$ commonly exhibit the FM phase with large values of $M_s$ and finite $H_c$.

To further elucidate the $V$-induced emergence of the FM phase, we investigated the electronic structure of the defect-engineered ETO thin films. Figure 4(a) shows the optical absorption coefficient spectra as functions of photon energy, $\alpha(\omega)$, for ETO thin films on LSAT substrates. $\alpha(\omega)$ was calculated using Beer's law, $\log(1/Tr) = \alpha t$, where $t$ is the thickness of the film and $Tr$ (= $Tr_{film//sub}/Tr_{sub}$) is the intensity ratio between the transmittance of the film//substrate ($Tr_{film//sub}$) and the substrate alone ($Tr_{sub}$). The $\alpha(\omega)$ of all the ETO thin films exhibits two peaks, i.e., one broad peak centered at $\sim 2.5$ eV and a sharp peak at $> 4.5$ eV. The $\sim 2.5$ eV and $> 4.5$ eV peaks correspond to the electronic transitions from the occupied Eu $4f$ and O $2p$ states to the unoccupied Ti 3$d$ state, respectively, as schematically shown in the inset of Fig. 4(a)[25]. To estimate the interband transition energies of ETO, $E_1$ and $E_2$ in ETO thin films on LSAT substrates were determined from linear extrapolations of $(\alpha\omega)^2$ using Tauc's law, as shown in Fig. 4(b). The results indicate direct optical gaps[51]. $E_1$ and $E_2$ can be defined as the transition energies from the tops of the occupied Eu 4$f$ and O 2$p$ bands to the bottom of the unoccupied Ti 3$d$ band, respectively.

A distinct difference in $\alpha(\omega)$ can be found between the FM and non-FM phase ETO thin films. In particular, the first peak is suppressed and broadened for the as-grown ETO thin films fabricated at high $P(O_2)$ (green and yellow solid lines; small $V$, non-FM), which have a smaller bandgap of $E_1$. In contrast, the as-grown ETO thin films grown at low $P(O_2)$ and the post-annealed thin films (red/blue solid and red/green dashed lines; large $V$, FM) exhibit a distinctive peak at ~2.5 eV with enhanced values of $E_1$. Since the Eu 4$f$ state is strongly localized just beneath the Fermi level ($E_F$), the characteristics of the first peak primarily depend on the unoccupied Ti 3$d$ state (see Fig. 1). As $V$ increases, the Ti 3$d$ state becomes more localized and hence narrower, leading to a more distinct peak at ~2.5 eV. This would increase the bandgap of $E_1$ as well[21], consistent with the experimental observations. On the other hand, the O 2$p$ state moves up toward $E_F$ as $V$ increases[21]. This would induce a red-shift of the second peak and decrease $E_2$ for the as-grown ETO thin films grown at low $P(O_2)$ and the post-annealed thin films, as shown in Fig. 4(a). Figure 4(c) summarizes the $E_1$ and $E_2$ values. Low $P(O_2)$ growth and/or post-annealing (large $V$) result in larger $E_1$ and smaller $E_2$ compared to the high $P(O_2)$ growth. Such changes in the electronic structure lead to the FM phase (blue box region).

Figures 5(a)-5(c) summarizes the experimental and theoretical results of the defect engineering in ETO thin films, which can be scaled using a single parameter $\Delta V$. Figure 5(a) shows $M_s$ as a function of $\Delta V$ for the ETO thin films extracted from Fig. 2 (films on LSAT substrates) and Fig. S2 (films on STO substrates), where $\Delta V$ (%) = 100 × $(V_{film} - V_{bulk})/V_{bulk}$ with $V_{bulk}$ = $(3.905)^3$ Å$^3$ = 59.547 Å$^3$. The FM phase with significantly large $M_s$ and finite $H_c$ emerges as $\Delta V$ increases, independent of the substrates used. Figure 5(b), replotted from Fig. 4(c), shows the systematic increase of $E_1$ and decrease of $E_2$ with $\Delta V$. The DFT calculation results consistently support the experimental observations, as shown in Fig. 5(c), although the quantitative values are rather different, possibly because of the typical underestimation of the DFT bandgap.

The consideration of the defects provides further insight into the origin of the FM phase in ETO. According to the Goodenough-Kanamori-Anderson rules, the super-exchange coupling between the next-next-nearest neighbor ($J_3 < 0$) depends strongly on the bond angle. In ETO, the Eu-Ti-Eu super-exchange interaction, which is known to stabilize the AFM state in bulk, is suppressed by the reduction of the Eu-Ti-Eu bond angle from 180°. Indeed, a similar interpretation has also been qualitatively given for the FM ground state of vertically strained ETO thin films[50]. To validate the scenario, we calculated the Eu-Ti-Eu bond angle in the vacant ETO. To validate this scenario, we calculated the Eu-Ti-Eu bond angle in the defect-engineered ETO. As shown in Fig. 5(d), the calculated Eu-Ti-Eu bond angle near the Eu and O vacancies in both the high- (cubic) and low-temperature (tetragonal) phases deviates from the initial value of 180° by 2–4°. These values are larger than that of ~1° expected in the vertically strained ETO thin films mentioned earlier[50]. Hence, the reduced Eu-Ti-Eu bond angle suppresses the AFM $J_3$ via Eu-Ti-Eu super-exchange in defect-engineered ETO thin films.

The nearest neighbor exchange coupling $J_1$ is another quantity that explains the emergent FM phase in the defect-engineered ETO thin films. The AFM super-exchange interaction between the Eu $4f$ and Ti $3d$ states is approximately represented as a third-order perturbation contribution, i.e., $J_1 \propto \sigma^4/\Delta E^2$, where $\Delta E$ and $\sigma$ are the bandgap and the overlap integral between the Eu $4f$ and Ti $3d$ states, respectively[52]. The overlap integral $\sigma$ can be estimated by calculating the integration of the crystal orbital Hamiltonian populations (ICOHP) values. As shown in Fig. S3, the formation of Eu and O vacancies by an enhanced $\Delta V$ increases the number of bonds with small Eu-Ti overlap (small |ICOHP| values) and reduces $\sigma$ (see purple symbols in Fig. S3). In contrast, the optical band gap $E_1$, responsible for $\Delta E$, increases with $\Delta V$ (Figs. 5 (b) and 5(c)), resulting in the suppression of the AFM $J_1$ at large $\Delta V$.

Last but not least, we consider the development of Eu-O-Eu super-exchange as another possible cause of the emergent FM ground state (Fig. 1). Defects break the surrounding local symmetry. The bond angles and lengths near the defect sites are hence often strongly distorted. In ETO, the Eu-O vacancy in conjunction with the enlarged $V$ may cause the Eu-O-Eu bond angle to deviate from 90°, which can in turn inhibit the AFM interaction or introduce new FM interactions. The strong Ti-O hybridization (> 4.5 eV peak in Fig. 4(a)) facilitates electron transfer from the occupied O $2p$ orbital to the unoccupied Ti $3d$ orbital. This introduces an FM exchange between the Eu ions via the partially occupied and spin-polarized O $2p$ orbital[53]. We note that a similar FM mechanism has been suggested in insulating $Cr_2MoO_6$, in which Cr-O-Cr super-exchange is facilitated by the strong Mo-O hybridization[53].

## 4. CONCLUSIONS

We studied the magnetic ground state of defect-engineered ETO epitaxial thin films. By controlling the $P(O_2)$ during growth and further applying post-annealing, we obtained ETO thin films with various $V$s ranging from $\Delta V = -3\%$ to 1%. A clear FM phase was observed in thin films with large $\Delta V$. Optical spectroscopy revealed that ETO thin films with large $\Delta V$ have larger bandgaps. The large band gap together with the decreased Eu-Ti orbital overlap and Eu-Ti-Eu bond angle obtained from the DFT calculations are expected to efficiently suppress the AFM Eu-Ti-Eu super-exchange interaction. On the other hand, the introduction of local structural symmetry breaking was suggested to introduce the FM Eu-O-Eu super-exchange. Our results provide a pathway to systematically understand and control the FM phase in defect-engineered epitaxial ETO thin films.

**AKNOWLEDGEMENTS**

This work was supported by the Basic Science Research Programs through the National Research Foundation of Korea (NRF-2021R1A2C2011340, NRF-2018R1D1A1B07042753 and NRF-2020K1A3A7A09077715).

**AUTHOR CONTRIBUTIONS**

D. S. and W. S. C. designed and analyzed the experiments. D. S. synthesized the thin films and characterized the structural information. I. K. and M. C. performed the theoretical calculations. S. S. and S. P. performed magnetic measurement and performed/analyzed XPS measurement. Y.-S. S. and J. H performed the optical measurement. M. C. and W. S. C. led the project. The manuscript was written through the contributions of all authors. All authors have given approval to the final version of the manuscript.

**REFERENCES**

1. Aggarwal S, Ramesh R. Point defect chemistry of metal oxide heterostructures. Annu Rev Mater Sci. 1998; 28:463–99.

2. Kalinin SV, Spaldin NA. Functional Ion Defects in Transition Metal Oxides. Science. 2013;341(858):858-9.

3. Tuller HL, Bishop SR. Point Defects in Oxides: Tailoring Materials Through Defect Engineering. Annu Rev Mater Res. 2011;41:369-98.

4. Li WW, Shi JL, Zhang KHL, MacManus-Driscoll JL. Defects in complex oxide thin films for electronics and energy applications: challenges and opportunities. Mater Horiz. 2020;7(11):2832-59.

5. Ahadi K, Galletti L, Li Y, Salmani-Rezaie S, Wu W, Stemmer S. Enhancing superconductivity in $SrTiO_3$ films with strain. Sci Adv. 2019;5:eaaw0120.

6. Haeni JH, Irvin P, Chang W, Uecker R, Reiche P, Li YL, et al. Room-temperature ferroelectricity in strained $SrTiO_3$. Nature. 2004;430:758–61.

7. Lee SA, Jeong H, Woo S, Hwang JY, Choi SY, Kim SD, et al. Phase transitions via selective elemental vacancy engineering in complex oxide thin films. Sci Rep. 2016;6:23649.

8. Kang KT, Seo HI, Kwon O, Lee K, Bae JS, Chu MW, et al. Ferroelectricity in $SrTiO_3$ epitaxial thin films via Sr-vacancy-induced tetragonality. Appl Surf Sci. 2020;499:143930.

9. Yoon H, Choi M, Lim TW, Kwon H, Ihm K, Kim JK, et al. Reversible phase modulation and hydrogen storage in multivalent $VO_2$ epitaxial thin films. Nat Mater. 2016;15(10):1113-9.

10. Lee D, Jeon BC, Baek SH, Yang SM, Shin YJ, Kim TH, et al. Active control of ferroelectric switching using defect-dipole engineering. Adv Mater. 2012;24(48):6490-5.

11. Damodaran AR, Breckenfeld E, Chen Z, Lee S, Martin LW. Enhancement of ferroelectric Curie temperature in $BaTiO_3$ films via strain-induced defect dipole alignment. Adv Mater. 2014;26(36):6341-7.

12. Lee SA, Oh S, Lee J, Hwang JY, Kim J, Park S, et al. Tuning electromagnetic properties of SrRuO$_3$ epitaxial thin films via atomic control of cation vacancies. Sci Rep. 2017;7(1):11583.

13. Lee SA, Oh S, Hwang J-Y, Choi M, Youn C, Kim JW, et al. Enhanced electrocatalytic activity via phase transitions in strongly correlated SrRuO$_3$ thin films. Energy Environ Sci. 2017;10(4):924-30.

14. Cheng S, Li M, Deng S, Bao S, Tang P, Duan W, et al. Manipulation of Magnetic Properties by Oxygen Vacancies in Multiferroic YMnO$_3$. Adv Funct Mater. 2016;26(21):3589-98.

15. Jeen H, Choi WS, Freeland JW, Ohta H, Jung CU, Lee HN. Topotactic phase transformation of the brownmillerite SrCoO$_{2.5}$ to the perovskite SrCoO$_{3-\delta}$. Adv Mater. 2013;25(27):3651-6.

16. Khare A, Shin D, Yoo TS, Kim M, Kang TD, Lee J, et al. Topotactic Metal-Insulator Transition in Epitaxial SrFeO$_x$ Thin Films. Adv Mater. 2017;29(37):1606566.

17. Muñoz A, de la Calle C, Alonso JA, Botta PM, Pardo V, Baldomir D, et al. Crystallographic and magnetic structure of SrCoO$_{2.5}$ brownmillerite: Neutron study coupled with band-structure calculations. Phys Rev B. 2008;78(5):054404.

18. Choi WS, Jeen H, Lee JH, Seo SS, Cooper VR, Rabe KM, et al. Reversal of the lattice structure in SrCoO$_x$ epitaxial thin films studied by real-time optical spectroscopy and first-principles calculations. Phys Rev Lett. 2013;111(9):097401.

19. Chien C-L, DeBenedetti S, Barros FDS. Magnetic properties of EuTiO$_3$, Eu$_2$TiO$_4$, and Eu$_3$Ti$_2$O$_7$. Phys Rev B. 1974;10(9):3913-22.

20. Katsufuji T, Takagi H. Coupling between magnetism and dielectric properties in quantum paraelectric EuTiO$_3$. Phys Rev B. 2001;64(5):054415.

21. Wang X, Zhen S, Min Y, Zhou P, Huang Y, Zhong C, et al. Influence of strain on optical properties of multiferroic EuTiO$_3$ film: A first-principles investigation. J Appl Phys. 2017;122(19):194102.

22. McGuire TR, Shafer MW, Joenk RJ, Alperin HA, Pickart SJ. Magnetic Structure of EuTiO$_3$. J Appl Phys. 1966;37(3):981-2.

23. Bussmann-Holder A, Roleder K, Stuhlhofer B, Logvenov G, Lazar I, Soszynski A, et al. Transparent EuTiO$_3$ films: a possible two-dimensional magneto-optical device. Sci Rep. 2017;7:40621.

24. Jiang K, Zhao R, Zhang P, Deng Q, Zhang J, Li W, et al. Strain and temperature dependent absorption spectra studies for identifying the phase structure and band gap of EuTiO$_3$ perovskite films. Phys Chem Chem Phys. 2015;17(47):31618-23.

25. Fennie CJ, Rabe KM. Magnetic and electric phase control in epitaxial EuTiO$_3$ from first principles. Phys Rev Lett. 2006;97(26):267602.

26. Lee JH, Fang L, Vlahos E, Ke X, Jung YW, Kourkoutis LF, et al. A strong ferroelectric ferromagnet created by means of spin-lattice coupling. Nature. 2010;466(7309):954-8.

27. Li CD, Zhao JL, Dong ZC, Zhong CG, Huang YY, Min Y, et al. Strain induced magnetic transitions and spin reorientations in quantum paraelectric EuTiO$_3$ material. J Magn Magn. 2015;382:193-201.

28. Lee JH, Ke X, Podraza NJ, Kourkoutis LF, Heeg T, Roeckerath M, et al. Optical band gap and magnetic properties of unstrained EuTiO$_3$ films. Appl Phys Lett. 2009;94(21):212509.

29. Ranjan R, Sadat Nabi H, Pentcheva R. Electronic structure and magnetism of EuTiO$_3$: a first-principles study. J Condens Matter Phys. 2007;19(40):406217.

30. Akamatsu H, Kumagai Y, Oba F, Fujita K, Murakami H, Tanaka K, et al. Antiferromagnetic superexchange via 3$d$ states of titanium in EuTiO$_3$ as seen from hybrid Hartree-Fock density functional calculations. Phys Rev B. 2011;83(21):214421.

31. Tanaka K, Fujita K, Maruyama Y, Kususe Y, Murakami H, Akamatsu H, et al. Ferromagnetism induced by lattice volume expansion and amorphization in $EuTiO_3$ thin films. J Mater Res. 2013;28(8):1031-41.

32. Katsufuji T, Tokura Y. Transport and magnetic properties of a ferromagnetic metal: $Eu_{1-x}R_xTiO_3$. Phys Rev B 1999;60:R15021.

33. Akahoshi D, Horie H, Sakai S, Saito T. Ferromagnetic behavior in mixed valence europium ($Eu^{2+}/Eu^{3+}$) oxide $EuTi_{1-x}M_xO_3$ ($M$ = $Al^{3+}$ and $Ga^{3+}$). Appl Phys Lett. 2013;103(17):172407.

34. Akahoshi D, Koshikawa S, Nagase T, Wada E, Nishina K, Kajihara R, et al. Magnetic phase diagram for the mixed-valence Eu oxide $EuTi_{1-x}Al_xO_3$ ($0 \leq x \leq 1$). Phys Rev B. 2017;96(18):184419.

35. Akahoshi D, Miyamoto G, Hayakawa Y, Saito T. The magnetic properties of the mixed valence Eu oxide $EuTi_{1-x}Sc_xO_3$ ($0 \leq x \leq 1$). J Solid State Chem. 2019;280:120985.

36. Kususe Y, Murakami H, Fujita K, Kakeya I, Suzuki M, Murai S, et al. Magnetic and transport properties of $EuTiO_3$ thin films doped with Nb. Jpn J Appl Phys. 2014;53(5S1):05FJ7.

37. Li L, Zhou H, Yan J, Mandrus D, Keppens V. Research Update: Magnetic phase diagram of $EuTi_{1-x}B_xO_3$ ($B$ = Zr, Nb). APL Mater. 2014;2(11):110701.

38. Ryan PJ, Kim JW, Birol T, Thompson P, Lee JH, Ke X, et al. Reversible control of magnetic interactions by electric field in a single-phase material. Nat Commun. 2013;4:1334.

39. Kugimiya K, Fujita K, Tanaka K, Hirao K. Preparation and magnetic properties of oxygen deficient $EuTiO_{3-\delta}$ thin films. J Magn Magn. 2007;310(2):2268-70.

40. Shimamoto K, Hatabayashi K, Hirose Y, Nakao S, Fukumura T, Hasegawa T. Full compensation of oxygen vacancies in EuTiO$_3$ (001) epitaxial thin film stabilized by a SrTiO$_3$ surface protection layer. Appl Phys Lett. 2013;102(4):042902.

41. Bussmann-Holder A, Köhler J, Kremer RK, Law JM. Relation between structural instabilities in EuTiO$_3$ and SrTiO$_3$. Phys Rev B. 2011;83(21).

42. Morozovska AN, Glinchuk MD, Behera RK, Zaulychny B, Deo CS, Eliseev EA. Ferroelectricity and ferromagnetism in EuTiO$_3$ nanowires. Phys Rev B. 2011;84:205403.

43. Kresse G, Hafner J. *Ab initio* molecular dynamics for liquid metals. Phys Rev B Condens Matter. 1993;47(1):558-61.

44. Perdew JP, Burke K, Ernzerhof M. Generalized Gradient Approximation Made Simple. Phys Rev Lett. 1996;77(18):1396.

45. Dudarev SL, Botton GA, Savrasov SY, Humphreys CJ, Sutton AP. Electron-energy-loss spectra and the structural stability of nickel oxide: An LSDA1U study. Phys Rev B. 1998;57(3):1505.

46. Yang Y, Ren W, Wang D, Bellaiche L. Understanding and revisiting properties of EuTiO$_3$ bulk material and films from first principles. Phys Rev Lett. 2012;109(26):267602.

47. Deringer VL, Tchougreeff AL, Dronskowski R. Crystal orbital Hamilton population (COHP) analysis as projected from plane-wave basis sets. J Phys Chem A. 2011;115(21):5461-6.

48. Shkabko A, Xu C, Meuffels P, Gunkel F, Dittmann R, Weidenkaff A, et al. Tuning cationic composition of La:EuTiO$_{3-\delta}$ films. APL Mater. 2013;1(5):052111.

49. Fujita K, Wakasugi N, Murai S, Zong Y, Tanaka K. High-quality antiferromagnetic EuTiO$_3$ epitaxial thin films on SrTiO$_3$ prepared by pulsed laser deposition and postannealing. Appl Phys Lett. 2009;94(6):062512.

50. Lin Y, Choi EM, Lu P, Sun X, Wu R, Yun C, et al. Vertical Strain-Driven Antiferromagnetic to Ferromagnetic Phase Transition in EuTiO$_3$ Nanocomposite Thin Films. ACS Appl Mater Interfaces. 2020;12(7):8513-21.

51. Pankove JI. Optical Processes in Semiconductors. Dover, New York: Courier Corporation; 1975.

52. Akamatsu H, Kumagai Y, Oba F, Fujita K, Tanaka K, Tanaka I. Strong Spin-Lattice Coupling Through Oxygen Octahedral Rotation in Divalent Europium Perovskites. Adv Funct Mater. 2013;23(15):1864-72.

53. Zhu M, Do D, Dela Cruz CR, Dun Z, Cheng JG, Goto H, et al. Ferromagnetic superexchange in insulating Cr$_2$MoO$_6$ by controlling orbital hybridization. Phys Rev B. 2015;92(9):094419.

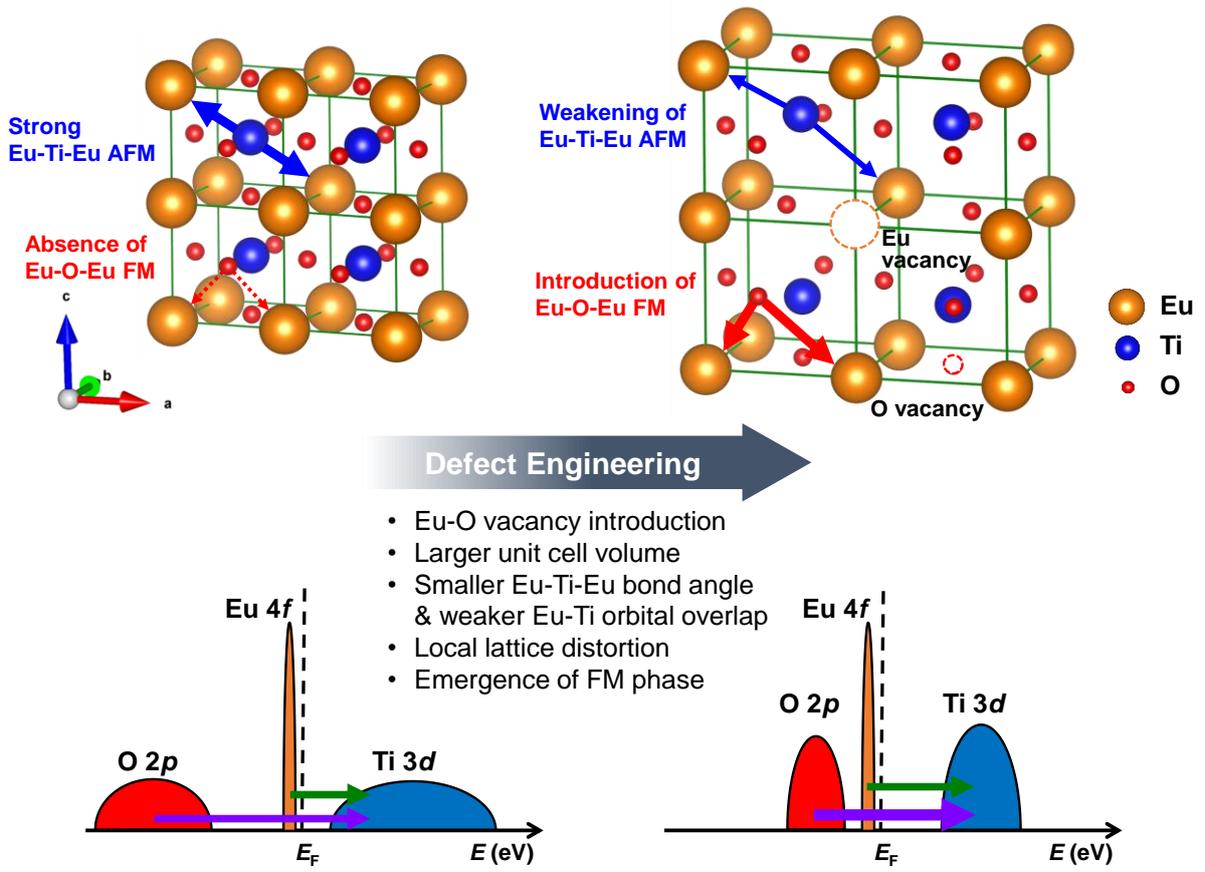

**FIGURE 1.**

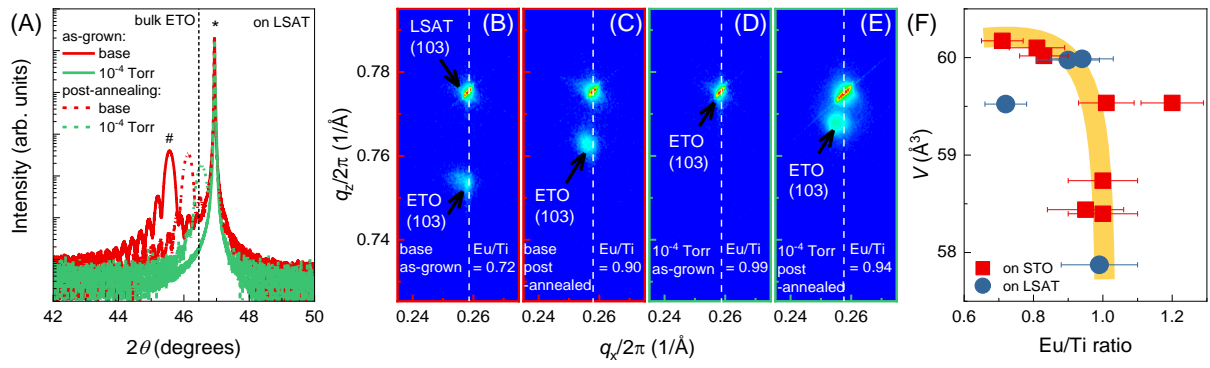

**FIGURE 2.**

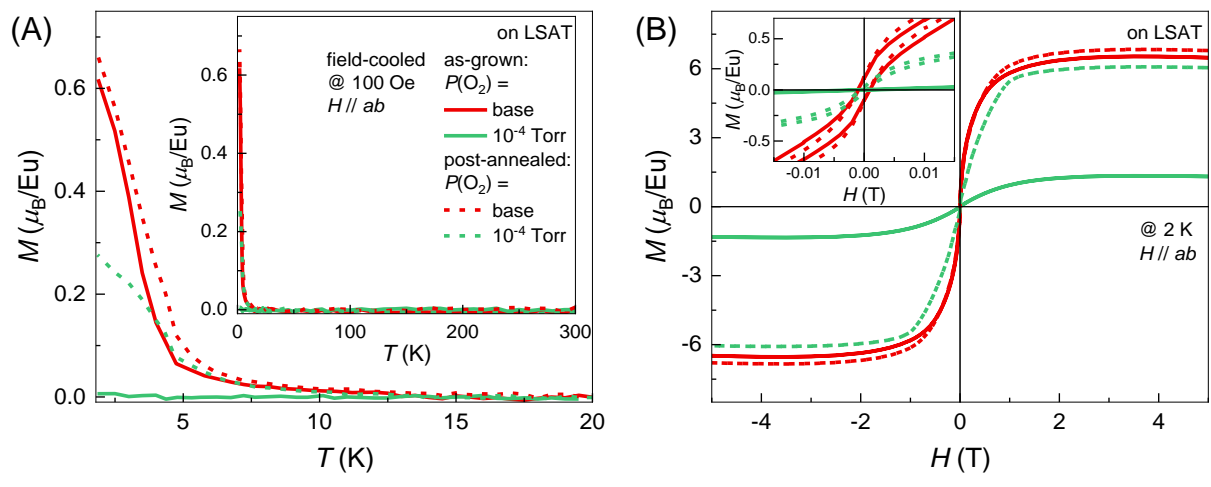

**FIGURE 3.**

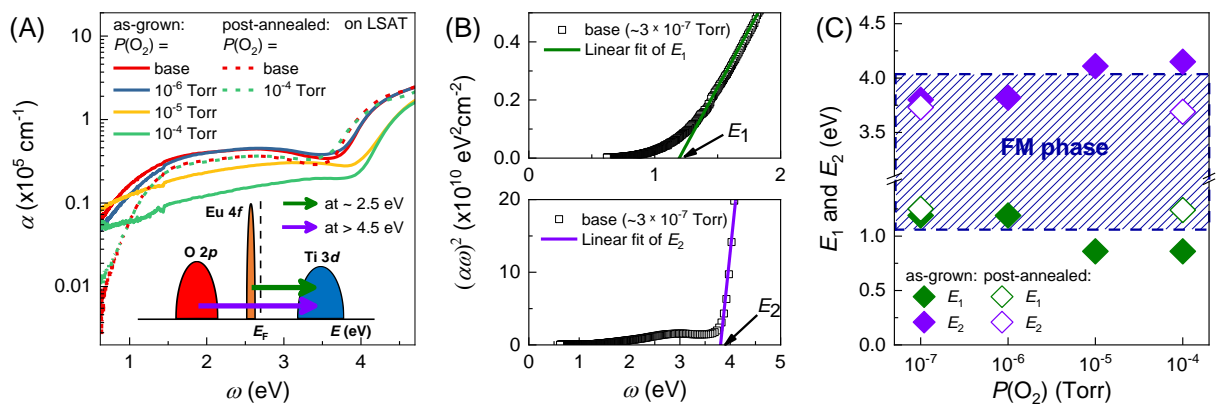

**FIGURE 4.**

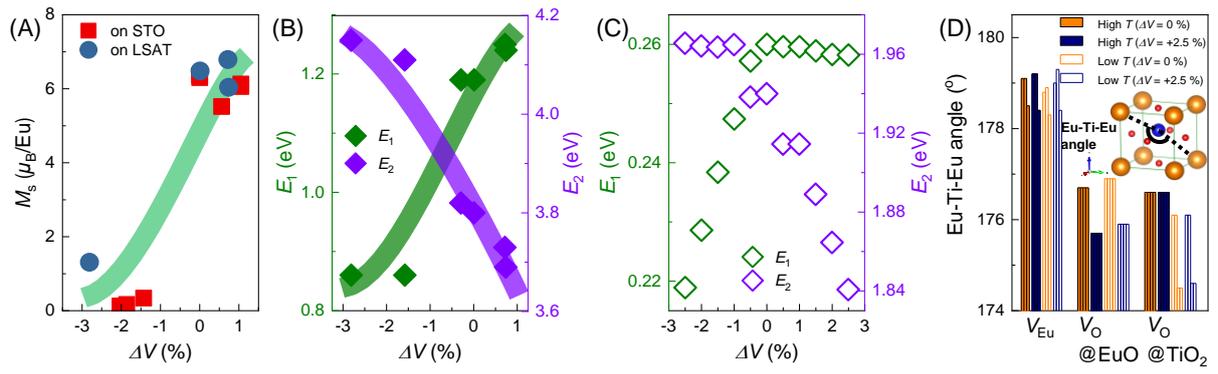

**FIGURE 5.**

**FIGURE CAPTIONS**

**FIGURE 1.** Schematic diagram of defect engineering in epitaxial ETO thin films. Defect engineering, especially the introduction of Eu-O vacancies (both Eu and O vacancies), results in u.c. volume expansion and lattice distortion, which lead to a smaller Eu-Ti-Eu bond angle, a weaker Eu-Ti orbital overlap, and local structural distortion. These effects collectively weaken the Eu-Ti-Eu AFM super-exchange interactions and give rise to the emergent FM phase (see the main text).

**FIGURE 2.** Crystal structures of epitaxial ETO thin films under various $P(O_2)$ and post-annealing. (A) XRD $\theta$-$2\theta$ scans near the (002) Bragg plane of for epitaxial ETO thin films (#) grown on LSAT substrates (*) at different $P(O_2)$ conditions ranging from the base pressure (~3 × $10^{-7}$ Torr) to $10^{-4}$ Torr. The solid (dotted) lines indicate the as-grown (post-annealed) thin films. Vertical dotted line indicates the $2\theta$ value of bulk ETO. (B)-(E) XRD reciprocal space maps (RSM) of the (B),(D) as-grown and (C),(E) post-annealed ETO thin films around the (103) Bragg reflection of the LSAT substrates. The films were grown at (B),(C) $P(O_2)$ = base (~3 × $10^{-7}$ Torr) and (D),(E) $P(O_2)$ = $10^{-4}$ Torr. (F) $V$ as a function of the Eu/Ti ratio obtained in ETO thin films grown on both LSAT and STO substrates. The thick solid line is a guide to the eye.

**FIGURE 3.** Emergence of FM phase in defect-engineered epitaxial ETO thin films. (A) $M(T)$ and (B) $M(H)$ curves of ETO thin films. (A) The $M(T)$ curves were obtained during field-cooled cooling at 100 Oe of the in-plane magnetic field. The inset shows a larger temperature range. (B) $M(H)$ curves obtained at 2 K. The inset shows a small-$H$ region in which the magnetic hysteresis loops are clearly visible.

**FIGURE 4.** Optical properties and electronic structures of the epitaxial ETO thin films. (A) $\alpha(\omega)$ of the ETO thin films. The inset schematically shows the band structures and optical transitions, where the olive and violet arrows correspond to the first peak at ~2.5 eV and the second peak at > 4.5 eV, respectively. (B) Linear extrapolation of $(\alpha\omega)^2$ based on Tauc's law to estimate the band gaps of $E_1$ (~1.0 eV) and $E_2$ (~4.0 eV) in the ETO thin films. (C) $P(O_2)$ and post-annealing-dependent $E_1$ and $E_2$. The blue dashed region indicates the FM phase.

**FIGURE 5.** Summary of the experimental ((A) and (B)) and theoretical ((C) and (D)) results for the magnetic properties and electronic structures as functions of $\Delta V$. (A) Saturation magnetization $M_s$ at 2 K and (B) band gaps as functions of $\Delta V$ extracted from Fig. 3(B) and Fig. 4(C), respectively. The thick solid lines are guides to the eye. (C) $\Delta V$ dependence of the band gaps ($E_1$ and $E_2$) extracted from DFT band structures. (D) The computed Eu-Ti-Eu angles in the presence of Eu and O vacancies.

# Supplementary Information

# Defect engineering of the magnetic phase in EuTiO$_3$ epitaxial thin films

*Dongwon Shin[1,#], Inseo Kim[2,#], Sehwan Song[3], Yu-Seong Seo[1], Jungseek Hwang[1], Sungkyun Park[3], Minseok Choi[2,\*], and Woo Seok Choi[1,\*]*

[1]*Department of Physics, Sungkyunkwan University, Suwon 16419, Republic of Korea*

[2]*Department of Physics, Inha University, Incheon 22212, Republic of Korea*

[3]*Department of Physics, Pusan National University, Busan 46241, Republic of Korea*

*Corresponding Authors: minseok.choi@inha.ac.kr; choiws@skku.edu

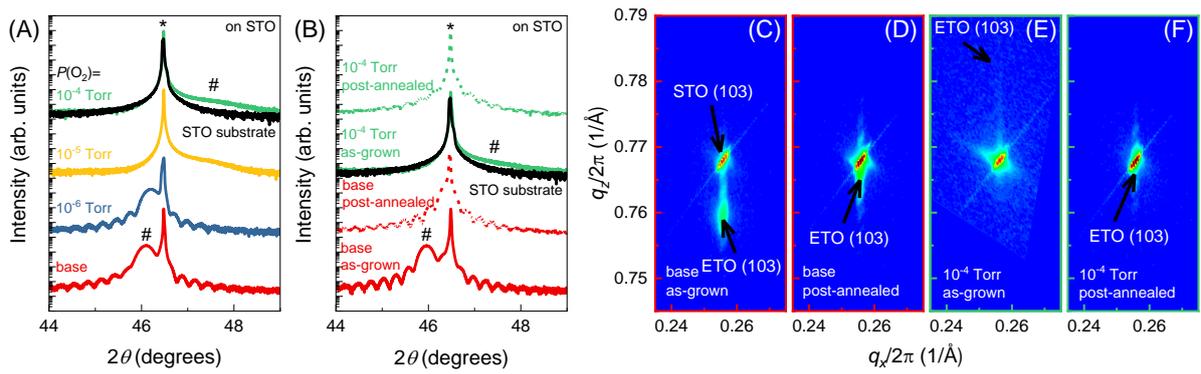

**Fig. S1.** Crystal structures of the epitaxial ETO thin films under various $P(O_2)$ and post-annealing. (A),(B) XRD $\theta$-$2\theta$ scans near the (002) Bragg plane of the epitaxial ETO thin films (#) on STO substrates (*) grown at different $P(O_2)$ conditions ranging from the base pressure ($\sim 3 \times 10^{-7}$ Torr) to $10^{-4}$ Torr. The solid (dotted) lines indicate the as-grown (post-annealed) thin films. The black solid lines denote the result of the STO substrate. (C)-(F) XRD RSM of the (C),(E) as-grown and (D),(F) post-annealed ETO thin films around the (103) Bragg reflection of the STO substrates. The films were grown at (C),(E) $P(O_2)$ = base ($\sim 3 \times 10^{-7}$ Torr) and (D),(F) $P(O_2)$ = $10^{-4}$ Torr.

| Substrate | $P(O_2)$ | Post-annealed | Eu (%) | Ti (%) | O (%) |
|---|---|---|---|---|---|
| LSAT | $10^{-4}$ Torr | X | 19.02 | 19.24 | 61.73 |
| LSAT | $10^{-4}$ Torr | O | 20.27 | 21.45 | 58.28 |
| LSAT | base (~ $3 \times 10^{-7}$ Torr) | X | 16.12 | 22.28 | 61.60 |
| LSAT | base (~ $3 \times 10^{-7}$ Torr) | O | 18.84 | 20.98 | 60.19 |
| STO | $10^{-4}$ Torr | X | 19.56 | 20.60 | 59.64 |
| STO | $10^{-4}$ Torr | O | 24.18 | 20.09 | 55.73 |
| STO | $10^{-5}$ Torr | X | 19.73 | 19.66 | 60.60 |
| STO | $10^{-6}$ Torr | X | 18.65 | 22.55 | 58.80 |
| STO | base (~ $3 \times 10^{-7}$ Torr) | X | 17.01 | 23.94 | 58.98 |
| STO | base (~ $3 \times 10^{-7}$ Torr) | O | 21.49 | 21.21 | 57.30 |

**Table S1.** Stoichiometry information of the ETO thin films extracted from XPS measurements

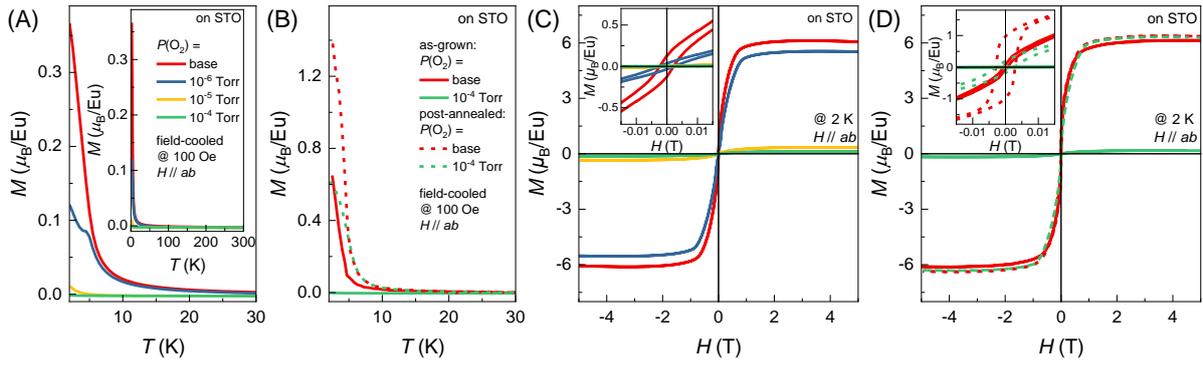

**Fig. S2.** Emergence of FM phase in defect-engineered epitaxial ETO thin films fabricated on STO substrates. (A),(B) $M(T)$ and (C),(D) $M(H)$ curves of the ETO thin films fabricated at different $P(O_2)$ conditions ranging from the base pressure ($\sim 3 \times 10^{-7}$ Torr) to $10^{-4}$ Torr. The solid (dotted) lines indicate the as-grown (post-annealed) thin films. (A),(B) The $M(T)$ curves were obtained during field-cooled cooling at 100 Oe of the in-plane magnetic field. The inset shows a larger temperature range. (C),(D) $M(H)$ curves were obtained at 2 K. The inset shows a small-$H$ region that in turn shows the magnetic hysteresis loops.

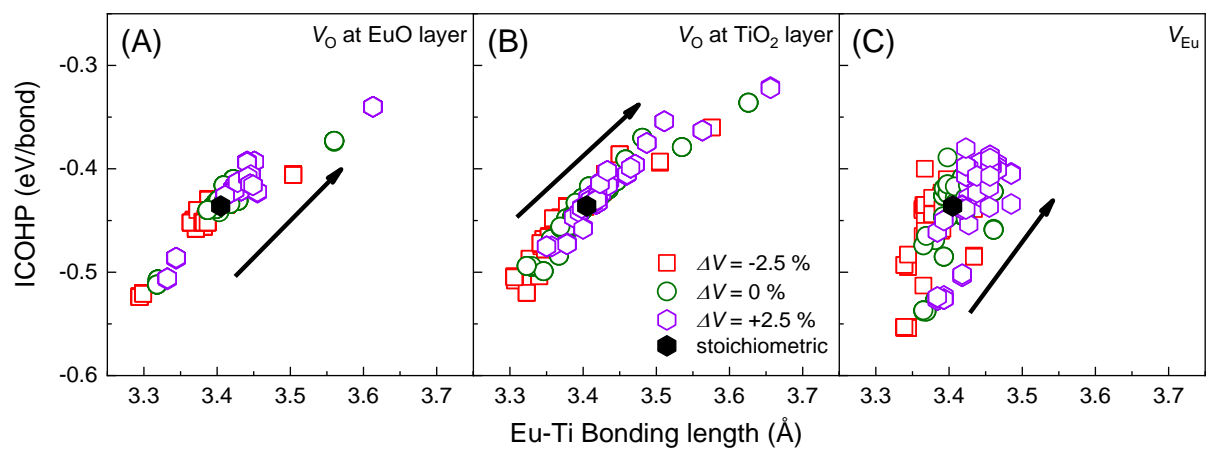

**Fig. S3.** (A)-(C) −ICOHP of between Eu and Ti as a function of the Eu-Ti bond length in ETO thin films with varying defects.

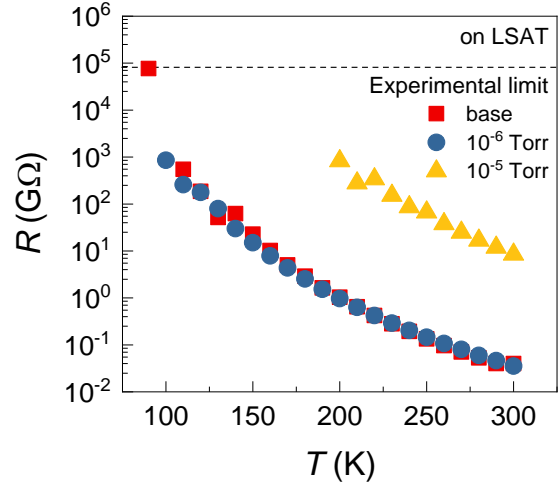

**Fig. S4.** Temperature-dependent electrical resistance of ETO thin films fabricated on LSAT substrate under different $P(O_2)$ conditions. Despite the possible charge transfer to the Ti $3d$ orbitals, our ETO thin films remains insulating. Although the exact origin of the insulating behavior is not clear at the moment, we speculate that the increased defect concentration in the thin film might have suppressed the conductivity. Nevertheless, the RKKY interaction adopted in many previous studies to explain the FM behavior in chemically-doped metallic ETO thin films cannot be applied to our samples.